\documentclass[%
 reprint,
 amsmath,amssymb,
 aps,
]{revtex4-2}

\usepackage{graphicx}
\usepackage{dcolumn}
\usepackage{bm}
\usepackage{xcolor}
\usepackage{verbatim}
\usepackage{ulem}
\usepackage{subfiles}

\begin{document}

\preprint{APS/123-QED}

\title{Non-Reciprocal Supercurrents in a Field-Free Graphene Josephson Triode}

\author
{John Chiles$^{1\dagger}$*, Ethan G. Arnault$^{1\dagger}$, Chun-Chia Chen$^1$,  Trevyn F.Q. Larson$^1$,\\ Lingfei Zhao$^1$, Kenji Watanabe$^2$, Takashi Taniguchi$^2$,\\ Fran\c{c}ois Amet$^3$, Gleb Finkelstein$^1$
\\
\normalsize{$^{1}$Department of Physics, Duke University, Durham, NC 27701, USA}\\
\normalsize{$^{2}$National Institute for Materials Science, Tsukuba, 305-0044, Japan}\\
\normalsize{$^3$Department of Physics and Astonomy, Appalachian State University, Boone, NC 28607, USA}\\
\normalsize{$^\dagger$ These authors contributed to this work equally}\\
\normalsize{$^\ast$To whom correspondence should be addressed; E-mail:  john.chiles@duke.edu}}

\date{\today}

\begin{abstract}
Superconducting diodes are proposed non-reciprocal circuit elements that should exhibit non-dissipative transport in one direction while being resistive in the opposite direction. Multiple examples of such devices have emerged in the past couple of years, however their efficiency is typically limited, and most of them require magnetic field to function. Here we present a device achieving efficiencies upwards of 90\% while operating at zero field. Our samples consist of a network of three graphene Josephson junctions linked by a common superconducting island, to which we refer as a Josephson triode. The triode is tuned by applying a control current to one of the contacts, thereby breaking the time-reversal symmetry of the current flow. The triode's utility is demonstrated by rectifying a small (tens of nA amplitude) applied square wave. We speculate that devices of this type could be realistically employed in the modern quantum circuits. 

\end{abstract}

\maketitle


\section{Introduction}

Diodes form one of the most important building blocks in electronic circuits, since they can be used in AC-DC conversion, signal rectification, and photodetection. The utility of diodes resides in their ability to offer non-reciprocity -- a low resistance to current flowing in one direction and a high resistance to current flowing in the opposite direction. 
While traditional diodes exploit P-N interfaces in semiconducting materials, a flurry of theoretical interest has focused on developing their superconducting analogues~\cite{zhang2022general,Yuan2022,He2022,Daido2022,Ili2022,Halterman2022,Scammell2022,davydova2022}. 

While these studies have followed a decade-long explorations of non-reciprocal supercurrents~\cite{Reynoso2008,Zazunov2009,Reynoso2012,Yokoyama2013,Brunetti2013,Silaev2014,Yokoyama2014,Dolcini2015,Hoshino2018,Wakatsuki2018,Chen2018}, the recent interest is driven by the search of novel materials which break both the inversion and time-reversal symmetry, thereby intrinsically enabling the superconducting diode effect (SDE). Such materials have indeed been experimentally identified and investigated, ranging from metallic films and proximitized semiconductors to van der Waals heterostructures ~\cite{Wakatsuki2017,Ando2020,diezmerida2021,shin2021magnetic,Baumgartner2021,lin2022zerofield,Wu2022,Strambini2022,Narita2022,Bauriedl2022,Pal2022}. 
While this direction offers a probe into the symmetry properties of novel materials, the resulting devices typically have limited diode efficiency, which is defined as a ratio of supercurrent in the forward and backward directions. 

In the meanwhile, it has been realized that higher superconducting diode efficiency can be achieved in properly designed nanostructures with an external magnetic field applied to break the time-reversal symmetry~\cite{Lyu2021,Souto2022,Golod2022,Gupta2022,Fominov2022}.
Unfortunately, magnetic field is often undesired for integrating the diodes in superconducting circuits. In this work, we rectify this problem by creating superconducting diodes which can operate at zero external magnetic field and achieve efficiency approaching 100\%. Our devices are based on multiterminal Josephson junctions made in graphene. In the past few years, the multiterminal junctions have been realized in a variety of materials \cite{Strambini2016,Draelos2019,Graziano2020,Pankratova2020,Arnault2022,Graziano2022} and have even 
found a foothold as a solution to technological problems \cite{Uri2016, Lee2020}.

We utilize the developments of multi-terminal Josephson junction design to eliminate the necessity of an applied magnetic field to achieve the SDE. The structure is based on three graphene Josephson junctions tied at the central superconducting island (Fig. 1a). Without applying a DC offset bias, 
all junctions are superconducting and the $I-V$ curves are symmetric. By applying a dissipationless control current in one of the junctions, we break the time-reversal symmetry and tune the $I-V$ curves of the other two junctions, achieving the SDE efficiencies exceeding 90\%. Our devices are further tunable by electrostatic gating \cite{suppinfo}, which allows us to adjust the scale the supercurrent that can be rectified.

\begin{figure*}[htp]
    \centering
    \includegraphics[width=1.8\columnwidth]{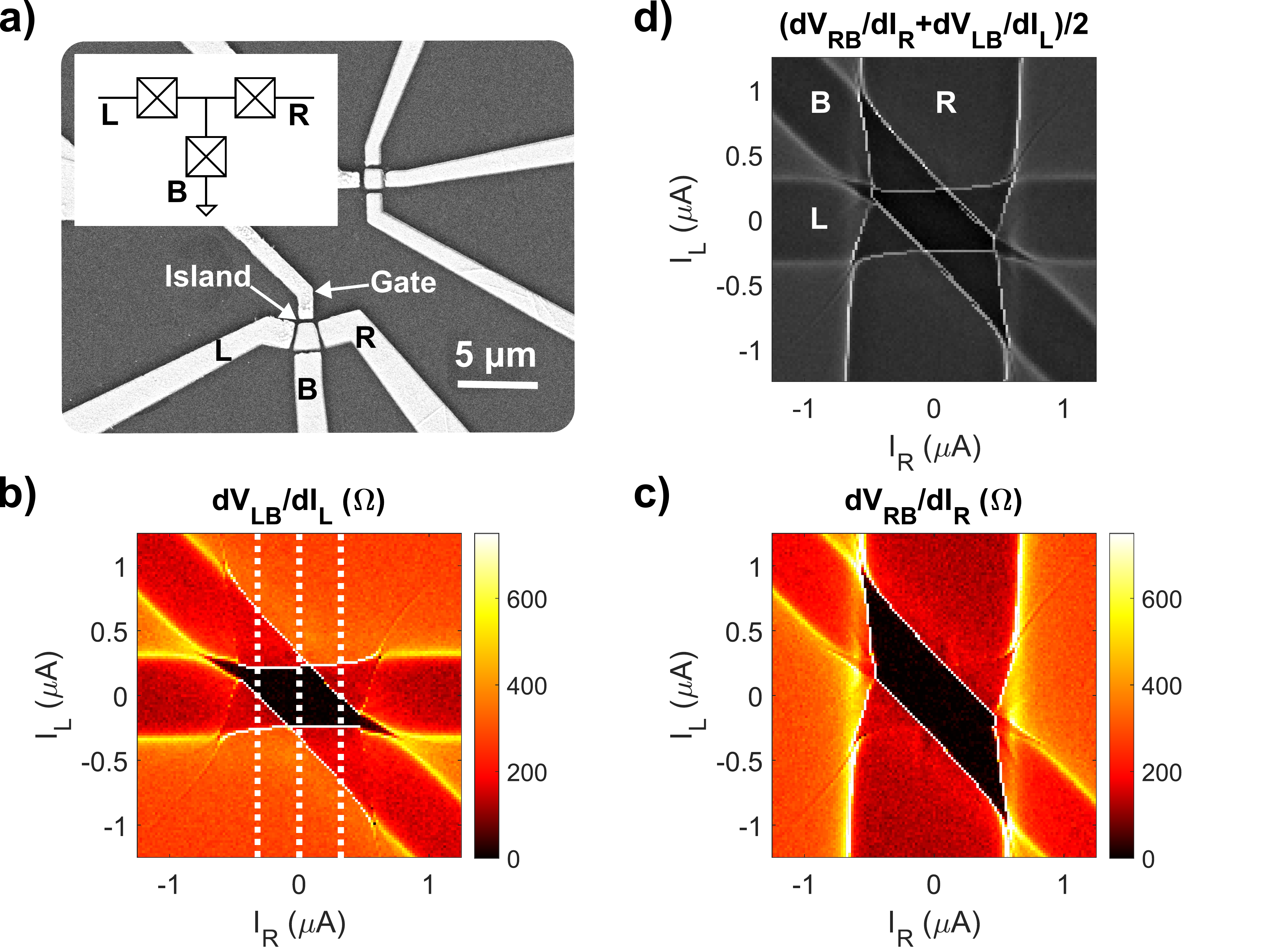}
    \caption {a) An SEM image of the device and a schematic of the relevant Josephson network (inset). We measure the bottom one of the two similar devices, which is comprized of a superconducting island connected to three superconducting leads (L, R, B) via graphene junctions (not visible). The gate electrode approaching the island from the top is not used here. b) Map of $\frac{dV_{LB}}{dI_{L}}$ with a central diamond corresponding to junctions L and B both superconducting. The dashed lines mark the locations of the $I-V$ cuts further analyzed in Fig. 2. c) A similar map of $\frac{dV_{RB}}{dI_{R}}$ with a central diamond corresponding to junctions R and B both superconducting. d) Grayscale enhanced-contrast map averaging the maps in (b) and (c). The black central region denotes the overlap of the central diamonds in (b) and (c) -- a roughly hexagonal region where all transport through the device is dissipationless.
    }
    \label{fig:Fig1}
\end{figure*}

\section{Results}

Our devices feature a trapezoidal superconducting island and three superconducting contacts labeled left (L), right (R) and bottom (B), Fig. 1a. (An additional side-gate approaching the island from the top does not contact the device and is not used in these measurements.) The contacts are connected to the island through 500 nm graphene channels which are etched such that none of the contacts are connected through graphene alone -- all transport must occur through the central island. We have measured two devices which yielded very similar results. For consistency, we present the results for one of them, shown at the bottom of Fig.~1a. 

The contacts and the island are made of sputtered molybdenum-rhenium (MoRe), which offers high-transparency Ohmic contact to graphene~\cite{Calado2015,Borzenets2016}. The devices are cooled to a base temperature of 60 mK in a Leiden Cryogenics dilution refrigerator.  A significant hysteresis is observed between the switching and retrapping currents, indicating either underdamping, or more likely electron overheating \cite{Borzenets2013}. To avoid this hysteresis, the sample is heated to $1.75 - 1.9$ K for the measurements in Figs. 1-3. We have verified that all the features measured at this temperature range exhibit negligible hysteresis and can be measured by sweeping the current in any direction. DC currents $I_{L}$ and $I_{R}$ are applied to the left and right contacts with respect to the grounded bottom contact, and a small AC excitation is used for extracting differential resistances. To maximize the critical current, a back gate voltage of 20 V is applied in Figs. 1, 2, and 4. The Dirac point in this sample is at 2.5 V. Importantly, all measurements take place at zero magnetic field. 

Fig. 1b and Fig. 1c correspond to the differential resistant maps measured between the left and bottom (1b) and right and bottom (1c) contacts. Correspondingly, the prominent horizontal (1b) and vertical (1c) features correspond to the regimes where respectively the left and right junctions are superconducting. The diagonal feature common to both maps corresponds to the bottom junction being superconducting. The device's collective behavior can be gleaned from Fig. 1d where $dV_{LB}/dI_{L}$ and $dV_{RB}/dI_{R}$ are averaged. The distinct black region in the center of this map occurs where all three junctions are superconducting and the transport across the entire device is dissipationless. 

\begin{figure*}[htp]
    \centering
    \includegraphics[width=2\columnwidth]{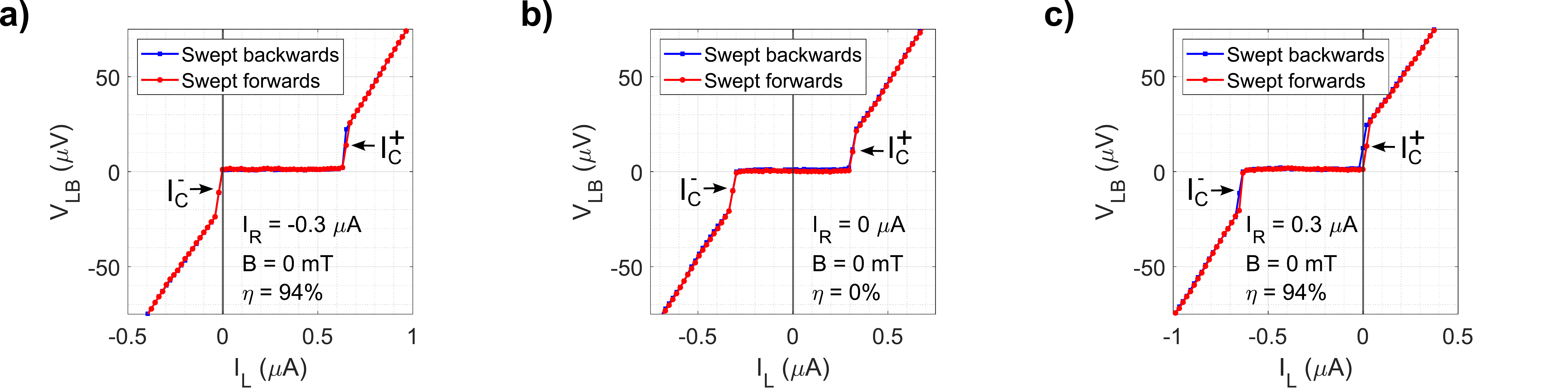}
    \caption {$I-V$ curves of the left junction measured at 3 values of $I_R$: a) $-0.3 \mu$A, b) $0 \mu$A, and c) $0.3 \mu$A. In all cases, the control current $I_R$ is well within the dissipationless range of the right junction. In all panels, $I_L$ is swept back and forth, showing negligible hysteresis. In panel (b), the $I-V$ curves are reciprocal, as expected. In (a) and (c), either $I_c^-$ or $I_c^+$ are approaching zero, resulting in a nearly ideal superconducting diode effect. 
        }
    \label{fig:Fig2}
\end{figure*}

The boundaries of the dissipationless region (an uneven hexagon) correspond to points where at least one junction on the device switches to normal. 
The non-reciprocity we exploit here appears in the regions of the hexagon defined by the switching currents of two junctions. An example is the left dashed line in Fig. 1b, which cuts through the hexagonal region away from its center. As a result, the system stays dissipationless only for positive $I_L$ and becomes dissipative at negative $I_L$ (Fig.~2a). Note that all three junctions play a role in this this process: the bottom junction switches at $I_L=I_c^- \approx 0$ while the left junction is responsible for the upper limit of the dissipationless range, $I_c^+$; finally, it is the biasing of the right junction which establishes the required asymmetry. For zero $I_R$, one recovers a symmetric cross-section along the $I_L$ direction (central white line in Fig.~1b and Fig.~2b). Finally, at the opposite value of the control current $I_R$ (right dash line in Fig.~1b and Fig.~2c) the system stays dissipationless only at negative $I_L \leq 0$ and the curves are reversed compared to Fig.~2a. Note that the right junction is biased below its critical current in all three cases. 

Collectively, the boundaries of the hexagon enable tuning of the transport non-reciprocity between any pair of contacts by adjusting the current applied to the third contact. This leads to a diode efficiency, \begin{equation} \eta = \frac{I_c^+ + I_c^-}{I_c^+ - I_c^-},\end{equation} 
that in practice can be tuned to exceed 90\%, as will be seen in Fig. 2, where we plot the $I-V$ curves corresponding to the three cross-sections in Fig. 1b. Each set of curves is measured in both directions, showing negligible hysteresis. When $I_{R} = 0$ (Fig.~2b) the curves are expectedly symmetric, so $I_c^+ = - I_c^-$ and $\eta=0$. However, as $I_{R}$ is tuned away from zero, the non-reciprocity grows and $|\eta|$ increases until it reaches $\pm 94\%$ at $I_R = \pm 0.3$ $\mu A$. 

Further increase of $\eta$ is possible by applying higher $I_R$. In fact, formally $\eta$ can exceed unity when $I_c^-$ becomes positive (same sign as $I_c^+$). However, this regime should be avoided if we are interested in rectifying small currents. Hence we stop increasing $|I_R|$ at the point where the high slope ``knee'' of the $I-V$ curve approaches the point $I_L=0$. We then define $I_c^-$ conservatively as a point at half the slope of the knee (see arrow in Fig. 2a), resulting in the $\eta \approx 94\%$. Finally, either positive (Fig. 2a) or negative (Fig. 2c) currents can be rectified depending on the desired operation.

\begin{figure}[htp]
    \centering
    \includegraphics[width=\columnwidth]{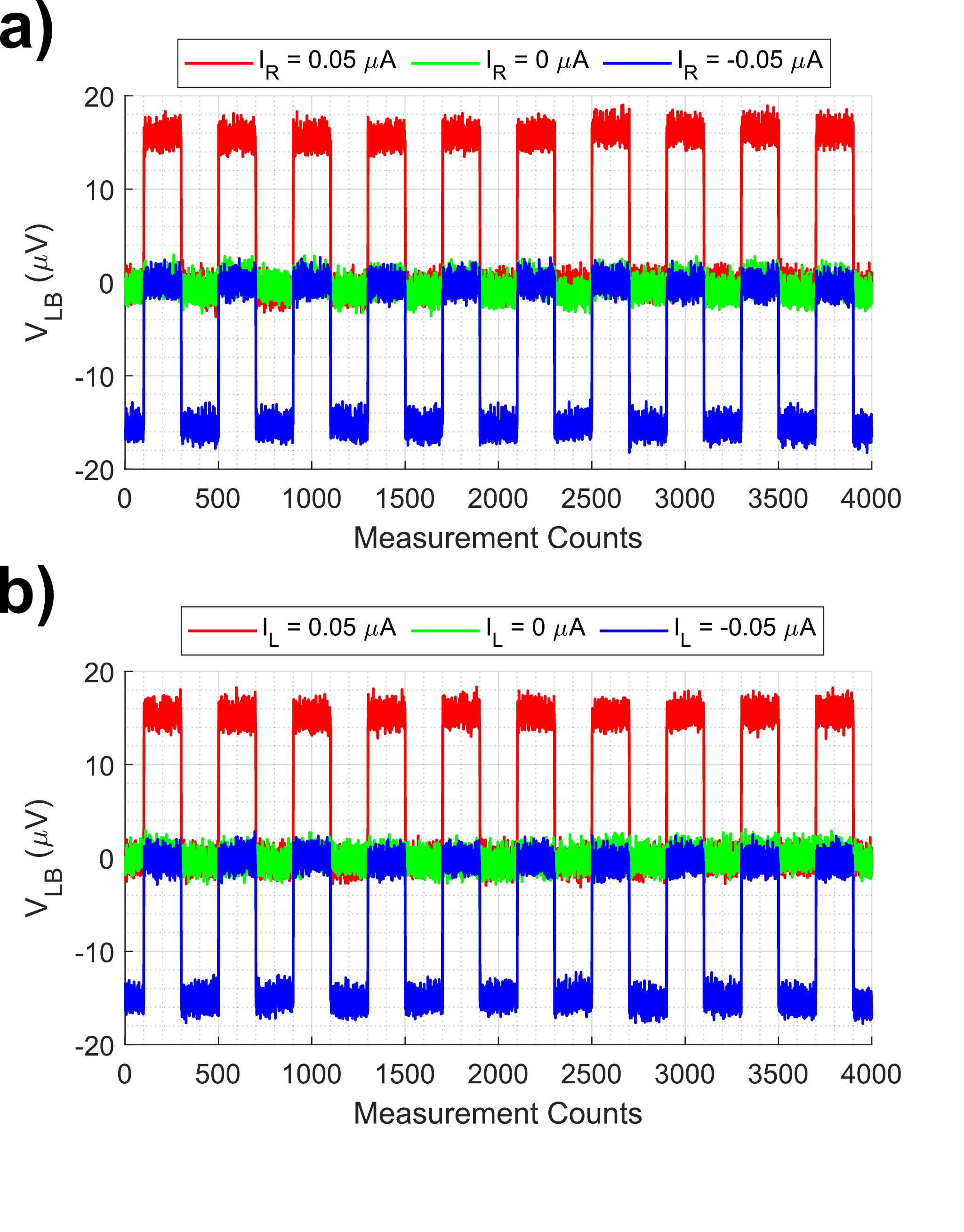}
    \caption {a) $V_{LB}$ measured as a square wave of a 60 nA amplitude is applied to $I_{L}$. Depending on the control current $I_R$, the negative and/or positive components of the square wave can be rectified (red and blue curve). For $I_R= 0$, all of the square wave lies within the dissipationless region. b) Now the square wave of amplitude $\pm$60 nA is applied to $I_{R}$ instead of $I_L$, while still monitoring $V_{LB}$. Depending on the value of $I_L$, the negative and/or positive components of the square wave can be rectified.
    }
    \label{fig:Fig3}
\end{figure}
To demonstrate the potential utility of the device in quantum circuits, we decrease the gate voltage to zero and apply a square wave of amplitude $\pm$ 60 nA to the left contact (Fig. 3a).  $I_{R}$ can then be set to produce desirable device responses: namely, at $I_{R} = 50$ and $-50$ nA,  the negative and positive portions of the square wave are respectively rectified. Further, when $I_{R}$ = 0, the device is fully superconducting for the entire square wave, as its amplitude is smaller than the critical current of the device. As a result, the entire square wave passes through the device without dissipation. Interestingly, we can change the biasing scheme to further utilize the triode's three-terminal nature. In Fig. 3b, we continue measuring $V_{LB}$ but now use $I_{L}$ as a control parameter, while applying a square wave to $I_{R}$. As a result, the voltage at the left contact, $V_{LB}$, switches depending on the sign of the square wave applied to the right contact. 

Fig.~4 explores the behavior of the sample as a function of temperature. The top row demonstrates the effect of temperature on the map of $R_{LB}$ vs $I_{L,R}$, first shown in Fig.~1. At the base temperature of 60 mK (Fig.~4a), the hysteresis is evident as the central superconducting pocket is shifted upward in the sweep direction. This hysteresis is still evident but much reduced at 1.3K (Fig.~4b), and finally it is clearly absent at 1.9 K (Fig.~4c). This is important, as hysteresis would prevent the diode from properly rectifying small currents. The bottom row presents the $I_L-V_{LB}$ curves, measured at the same three temperatures with the $I_R$ adjusted to optimally tune the diode efficiency. (The values of $I_R$ are indicated as white lines across the corresponding maps.) Again, pronounced hysteresis is observed at the lowest temperature, but it is almost gone at 1.3 K, while the upper switching current, $I_c^+ \approx 1 \mu$A, is not much suppressed compared to the lowest temperature. Further optimisation may result in suppressing the hysteresis at the base temperature. However, even the present device could be placed at the still plate of the dilution refrigerator to efficiently rectify a range of currents in the $0.1-1 \mu$A range. 

Interestingly, at base temperature, the central superconducting region deviates from the diamond present at elevated temperatures. In this state, nonlinear boundaries of the dissipationless region emerge -- boundaries where a small amount of additional current on one contact could switch the device while a larger change in current would be required on the other contact. Along such boundaries, true transistor-like amplification could be envisioned.

\begin{figure*}[htp]
    \centering
    \includegraphics[width=2\columnwidth]{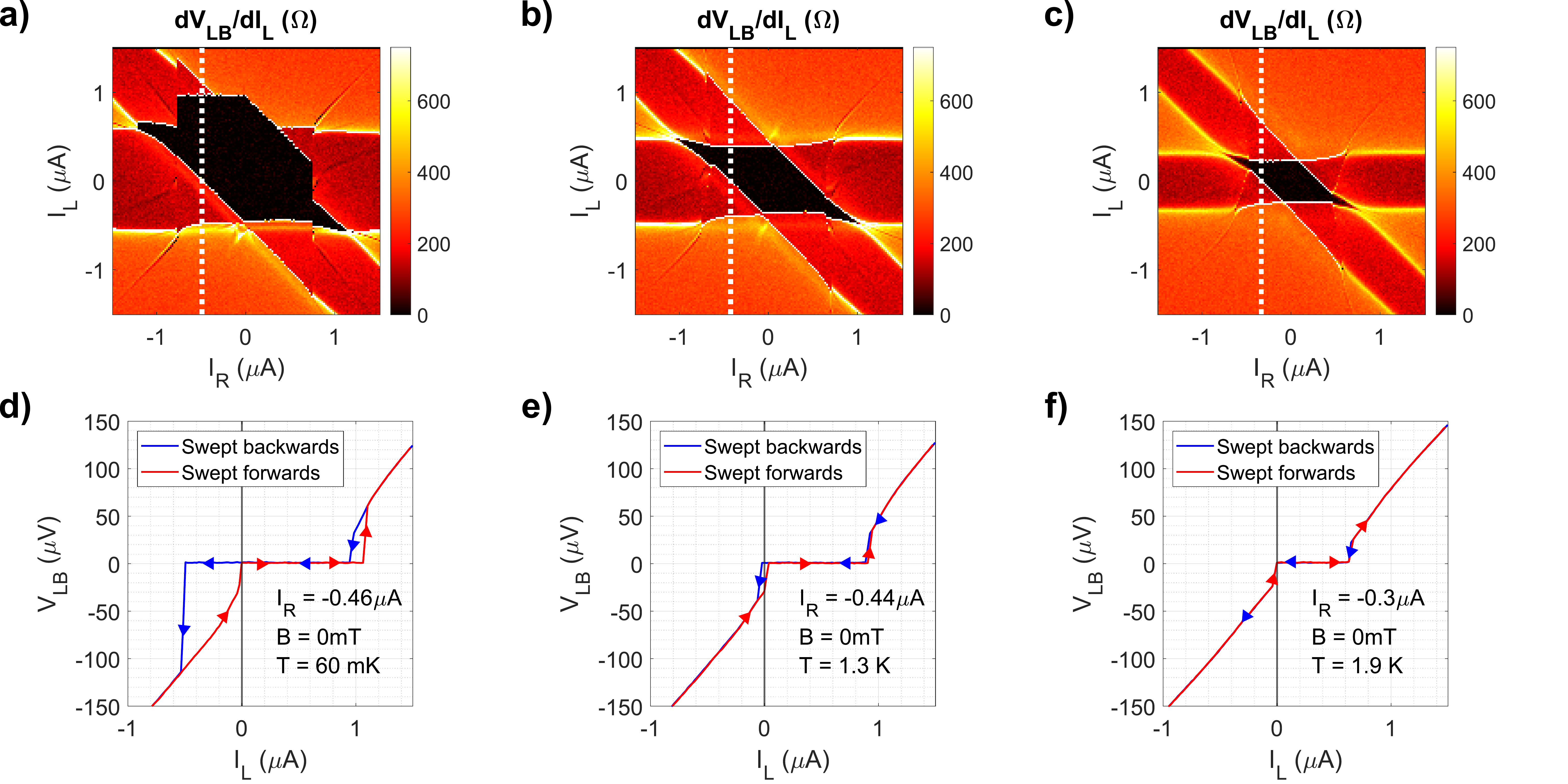}
    \caption {(a-c) Maps of $R_{LB}=\frac{dV_{LB}}{dI_{L}}$ taken at (a) 60 mK, (b) 1.3 K and (c) 1.9 K with $V_G=20$ V. As temperature increases, the substantial hysteresis present in panel (a) becomes negligible in the other two panels. (d-f) $I-V$ curves of the left junction measured at the same 3 temperatures. While the hysteresis disappears in panel (f), it is already small enough for practical purposes in panel (e). In all cases, the control current $I_R$ is adjusted for optimal diode performance and its values are indicated by dashed lines in panels (a-c). Panels (c) and (f) show the same data as Figs.~1b and 2a.
}
    \label{fig:Fig4}
\end{figure*}

\section{Discussion}

While the near perfect diode efficiency of our device is compelling, integration into quantum circuits may be fundamentally limited the frequency cutoff and the fluctuations in the diode efficiency.

The frequency cutoff is limited by the RC time constant of the device. In the present case, this can be estimated by taking the normal resistance of the graphene junction ($\sim$ 500 $\Omega$) and the capacitance of the bonding pads and leads to the silicon backgate ($\sim 2$ pF). Therefore, our present cutoff frequency could be as high as $\sim$ 1 GHz, indicating the promise for integrating our device into superconducting quantum circuits. Further improvements can be straightforwardly made by replacing the doped Si/SiO$_2$ substrate with undoped Si and moving to local gating schemes. This would greatly decrease the parasitic capacitance, especially when using the triode on-chip with the rest of the circuit.

Turning to the fluctuations of the diode efficiency, the noise of the DC control bias could be an important source. As this bias fluctuates, the diode will pick up slight changes of $I_c^-$, causing fluctuations of $\eta$. 
While in this measurement we have not optimized the current noise beyond reasonably acceptable levels, we have been able to limit it below $\sim 1$ nA in the past~\cite{Amet2016}. In principle the noise in the diode could be pushed down to the quantum limit, where it is dominated by critical current fluctuations~\cite{Haque2021}.

\section{Conclusion}
In summary, we utilize a network of three Josephson junctions at zero magnetic field to produce a superconducting diode with efficiency exceeding 90\%. The device geometry -- three graphene junctions connected by a common superconducting island -- provides high degree of control through applied currents and gate voltage. While this device is made fairly simply in graphene, one can imagine copying the same Josephson circuit in other materials, including Al/Al$_2$O$_3$ tunnel junctions. Although in that case the gate tunability would be lost, the advantage is that the fabrication techniques are well developed and easily scalable. Further effort in this direction could quickly present opportunities for  including high-efficiency superconducting diodes in quantum circuits.

\section{Methods}

The sample is fabricated using standard techniques in exfoliation, stacking, and electron-beam lithography. The device is etched with ChF$_{3}$O$_{2}$ and SF$_{6}$ plasma and contacts are deposited from sputtered molybdenum-rhenium (MoRe). 

Measurements were conducted in a Leiden Cryogenics dilution refrigerator at a base sample temperature of 60 mK. Two stage 1.5 k$\Omega$/1 nF RC filters are mounted at the mixing chamber. Current bias passed through 1 M$\Omega$ room temperature resistors, and voltage is measured with 1 nV$/\sqrt\textrm{Hz}$ $\times 100$ homemade amplifiers. Interfacing with computer is performed with an NI USB-6363 DAQ board for both current biasing and voltage readouts.

To produce differential resistance maps as in Figure 1, the current bias on one contact is stepped in individual steps in an outer loop, while the bias on the other contact is swept quickly. 
From here, numerical derivatives can easily be computed to give resistance values ($dV/dI$). Currents are ramped in both directions (e.g. $-1.5 \mu$A to 1.5 $\mu$A and back to $-1.5 \mu$A) and all derivatives are computed (e.g. $dV_{RB}/dI_{R}$ as well as $dV_{LB}/dI_{R}$) so several maps can be produced for each measurement.

\section{Acknowledgements}
We thank K.C. Fong for helpful discussions. Transport measurements by J.C., E.G.A., T.F.Q.L. and C.C., and data analysis by J.C., E.G.A., and G.F., were supported by Division of Materials Sciences and Engineering, Office of Basic Energy Sciences, U.S. Department of Energy, under Award No. DE-SC0002765. 
Sample fabrication and characterization by E.G.A. and L.Z. was supported by the NSF Award DMR-2004870. F.A. was supported by a URC grant at Appalachian State University. K.W. and T.T. acknowledge support from JSPS KAKENHI (Grant Numbers 19H05790, 20H00354 and 21H05233). The sample fabrication was performed in part at the Duke University Shared Materials Instrumentation Facility (SMIF), a member of the North Carolina Research Triangle Nanotechnology Network (RTNN), which is supported by the National Science Foundation (Grant ECCS-1542015) as part of the National Nanotechnology Coordinated Infrastructure (NNCI).

\newpage

\bibliography{apssamp}

\setcounter{figure}{0} 
\setcounter{section}{0}
\onecolumngrid
\newpage
\section{Supplementary Information}

\title{Supplementary Information for Non-Reciprocal Supercurrents in a Field-Free Graphene Josephson Triode}
\author
{John Chiles$^{1\dagger}$*, Ethan G. Arnault$^{1\dagger}$, Chun-Chia Chen$^1$,  Trevyn F.Q. Larson$^1$,\\ Lingfei Zhao$^1$, Kenji Watanabe$^2$, Takashi Taniguchi$^2$,\\ Fran\c{c}ois Amet$^3$, Gleb Finkelstein$^1$
\\
\normalsize{$^{1}$Department of Physics, Duke University, Durham, NC 27701, USA}\\
\normalsize{$^{2}$National Institute for Materials Science, Tsukuba, 305-0044, Japan}\\
\normalsize{$^3$Department of Physics and Astonomy, Appalachian State University, Boone, NC 28607, USA}\\
\normalsize{$^\dagger$ These authors contributed to this work equally}\\
\normalsize{$^\ast$To whom correspondence should be addressed; E-mail:  john.chiles@duke.edu}}

\maketitle
\subsection{Methods}
The sample is fabricated using standard techniques in exfoliation, stacking, and electron-beam lithography. The device is etched with ChF$_{3}$O$_{2}$ and SF$_{6}$ plasma and contacts are deposited from sputtered molybdenum-rhenium (MoRe).

Measurements were conducted in a Leiden Cryogenics dilution refrigerator at a base sample temperature of 60 mK. Two stage 1.5 k$\Omega$/1 nF RC filters are mounted at the mixing chamber. Current bias passed through 1 M$\Omega$ room temperature resistors, and voltage is measured with 1 nV$/\sqrt\textrm{Hz}$ $\times 100$ homemade amplifiers. Interfacing with computer is performed with an NI USB-6363 DAQ board for both current biasing and voltage readouts. 

To produce differential resistance maps as in Figure 1, the current bias on one contact is stepped in individual steps in an outer loop, while the bias on the other contact is swept quickly. 
From here, numerical derivatives can easily be computed to give resistance values ($dV/dI$). Currents are ramped in both directions (e.g. $-1.5 \mu$A to 1.5 $\mu$A and back to $-1.5 \mu$A) and all derivatives are computed (e.g. $dV_{RB}/dI_{R}$ as well as $dV_{LB}/dI_{R}$) so several maps can be produced for each measurement.

\subsection{Extended Temperature and Gate Voltage Data}
In Supplementary Figure 1, we present maps of $R_{RB}$ as $I_{R}$ and $I_{L}$ are varied. The maps are measured at 60 mK (a), 1.3 K (b), and 1.9 K (c), corresponding to the analogous $R_{LB}$ maps of in Figure 4 of the main text. The vertical dashed lines indicate the values of $I_R$ at which the $I - V$ cuts of Figure 4 were taken. It is clear that these values of are well below the critical current of the right junction, indicating dissipationless control. Hence, the whole triode operates in the dissipationless regime while $V_{LB} = 0$ for a given $I - V$ curve.

\begin{figure*}[htp]
    \centering
    \includegraphics[width=\columnwidth]{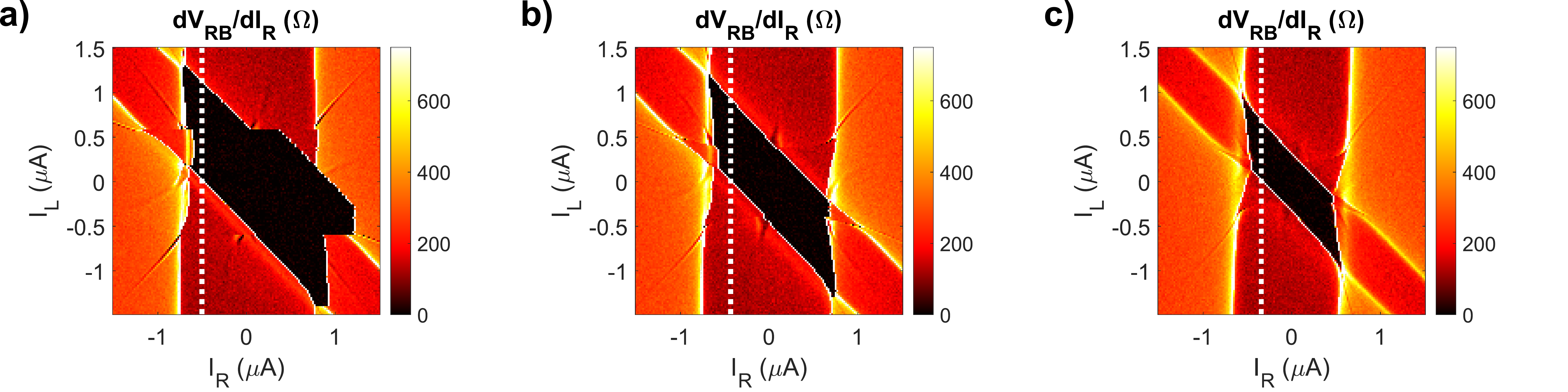}
    \caption {Maps of $\frac{dV_{RB}}{dI_{R}}$ measured at (a) 60 mK, (b) 1.3 K, and (c) 1.9 K. Dashed lines indicate the cross-sections where the $I - V$ curves are taken for the Figure 4d-e of the main text.
    }
    \label{fig:FigS1}
\end{figure*}

As initially presented in Figure 4 of the main text, the device can be operated at a range of applied gate voltages and temperatures.
To explore the full range of the operating parameters, we stepped the back gate voltage  between 0 V and 25 V in steps of 5 V, and stepped the sample temperature from 60 mK to 1.9 K in 6 steps. As a result, we have 72 bias-bias maps for both $\frac{dV_{LB}}{dI_{L}}$ and $\frac{dV_{RB}}{dI_{R}}$. In the Supplementary Figure 2, we present the $R_{LB}$ maps for $V_G=0$, 10, and 20 V. It is visible that the hysteresis disappears between 0.5 and 1K depending on the gate voltage, allowing for optimal operating conditions. Further, as gate is swept, we can rectify either smaller AC signals (near the Dirac peak at 2.5 V) or larger AC signals far from the Dirac peak. Notably, the high contact transparency of the MoRe-graphene interface allows for operating the device with both electron and hole doping.

\begin{figure*}[htp]
    \centering
    \includegraphics[width=0.8\columnwidth]{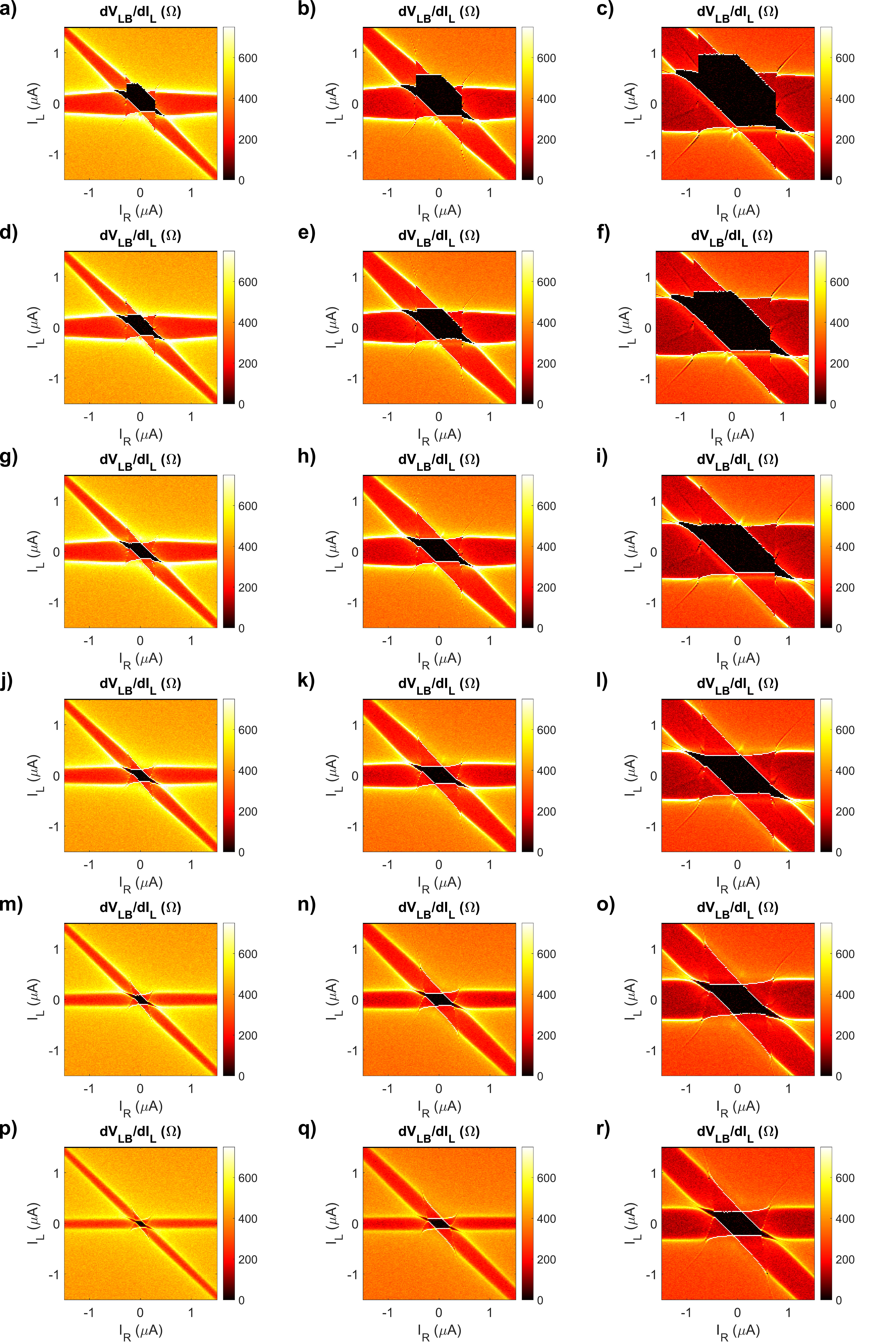}
    \caption {Maps of $\frac{dV_{LB}}{dI_{L}}$ measured with 0 V (left column), 10 V (middle column), and 20 V (right column) applied to the back gate. These maps are taken as the temperature is increased from top to bottom: (a-c) 60 mK,  (d-f) 530 mK, (g-i) 930 mK, (j-l) 1.3 K, (m-o) 1.6 K, and (p-r) 1.9 K.
    }
    \label{fig:FigS2}
\end{figure*}

\newpage
\subsection{Behavior in Applied Magnetic Field}

Supplemental Figure 3 shows the prototypical Fraunhofer patterns measure in perpendicular magnetic field for (a) $R_{LB}$ at $I_R=0$ and (b) $R_{RB}$ at $I_L=0$. The patterns emerge as expected with the maximal switching current appearing at zero field. 
As expected, multiple lobes of the pattern are observed. As the junctions areas slightly vary, so do their field periodicities. 
Notably, in the second lobe there is a reduced overlap of the Fraunhofer patterns, causing a larger dark orange region where one of the junctions is turned normal by the field. Again, this feature is expected as slight offsets in the periodicity of the Fraunhofer pattern will compound at higher magnetic fields. This mismatch in periodicity may likely be the factor limiting the device performance at elevated field.

\begin{figure*}[htp]
    \centering
    \includegraphics[width=\columnwidth]{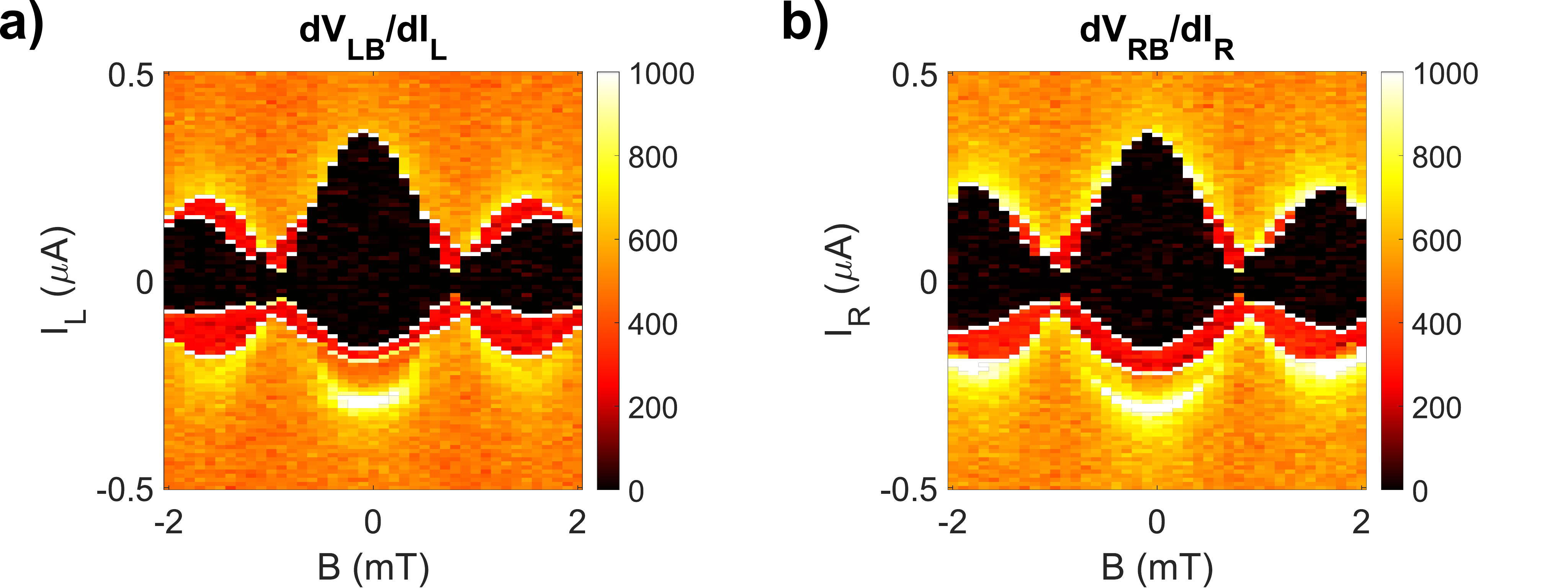}
    \caption {Fraunhofer patterns of (a) $R_{LB}=\frac{dV_{LB}}{dI_{L}}$ and (b) $R_{RB}=\frac{dV_{RB}}{dI_{R}}$ measured in the $\pm$2 mT range. All junctions exhibit similar critical currents and periodicity in field. These maps are taken at 60 mK with 0 V applied to the gate. $I_{R}$ and $I_{L}$ are fixed at 0 in (a) and (b), respectively, corresponding to a vertical and a horizontal cross-sections of the $I_{R}-I_{L}$ maps shown elsewhere in the text.
    }
    \label{fig:FigS3}
\end{figure*}

In Supplementary Figure 4, we show the Fraunhofer pattern with increased resolution. 
Note the region in both maps near 0.8 mT, where the dark orange corresponds to the bottom junction switching before either the left or right junction.
Also, the current maxima are shifted from the nominal zero field by $0.1$ mT = $1$ Gs, which is within the expected range of the systematic error of the power supply. When measuring the data for the rest of the paper, we applied a nominal $-0.1$ mT to stay at the true zero of magnetic field. 

\begin{figure*}[htp]
    \centering
    \includegraphics[width=\columnwidth]{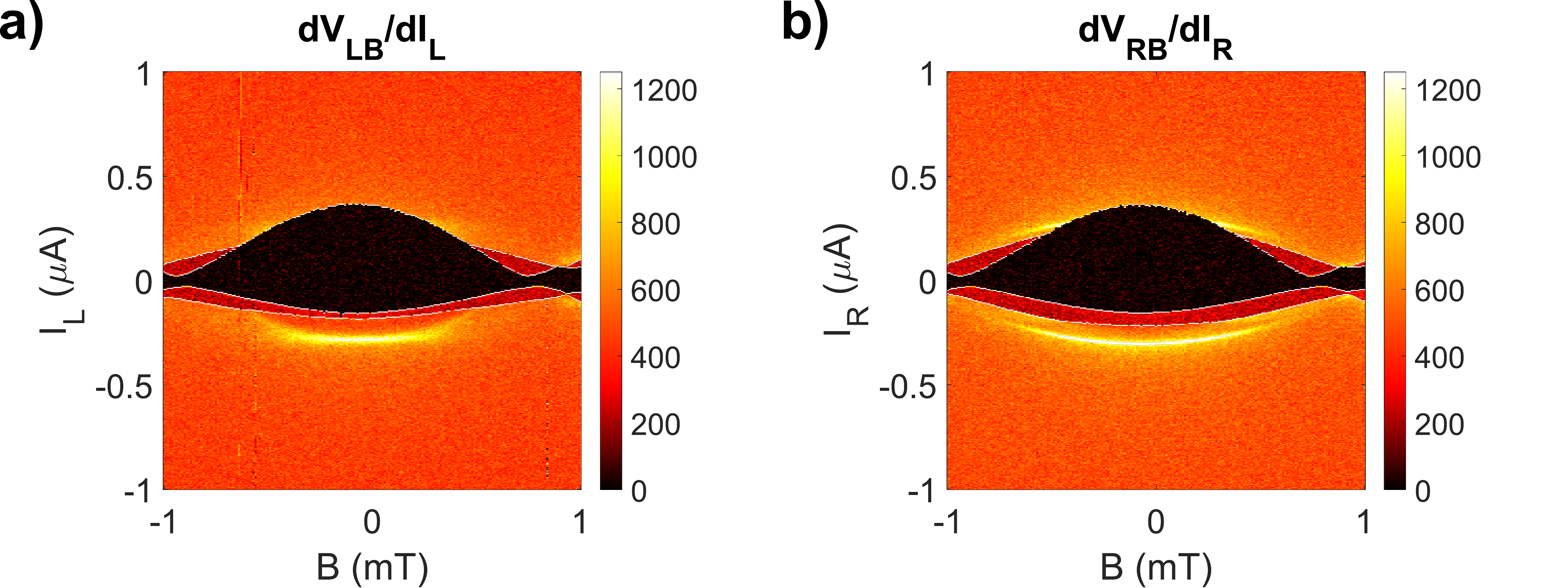}
    \caption {Higher resolution versions of the maps presented in Supplementary Figure 3, measured with the same temperature, gating, and biasing conditions. These maps are used to determine the true zero of the field which is used for the measurements in the rest of the paper. The apparent zero field is offset by 0.1 mT from the true zero due to the small systematic error of the magnet power supply.
    }
    \label{fig:FigS4}
\end{figure*}

While the triode can be operated without the application of magnetic field, it may occur that the device has to be placed in an environment with some magnetic field offset. To this end, we show its performance at a small perpendicular field. In Supplementary Figure 5, we demonstrate a square wave rectification of the AC signal at 1 K and an applied field of 0.6 mT. Even at this elevated field, we are able to rectify ambipolar AC signals for a wide range of current amplitudes, thus demonstrating the device's capability to work in the presence of a remnant or stray field.

\begin{figure*}[htp]
    \centering
    \includegraphics[width=0.5\columnwidth]{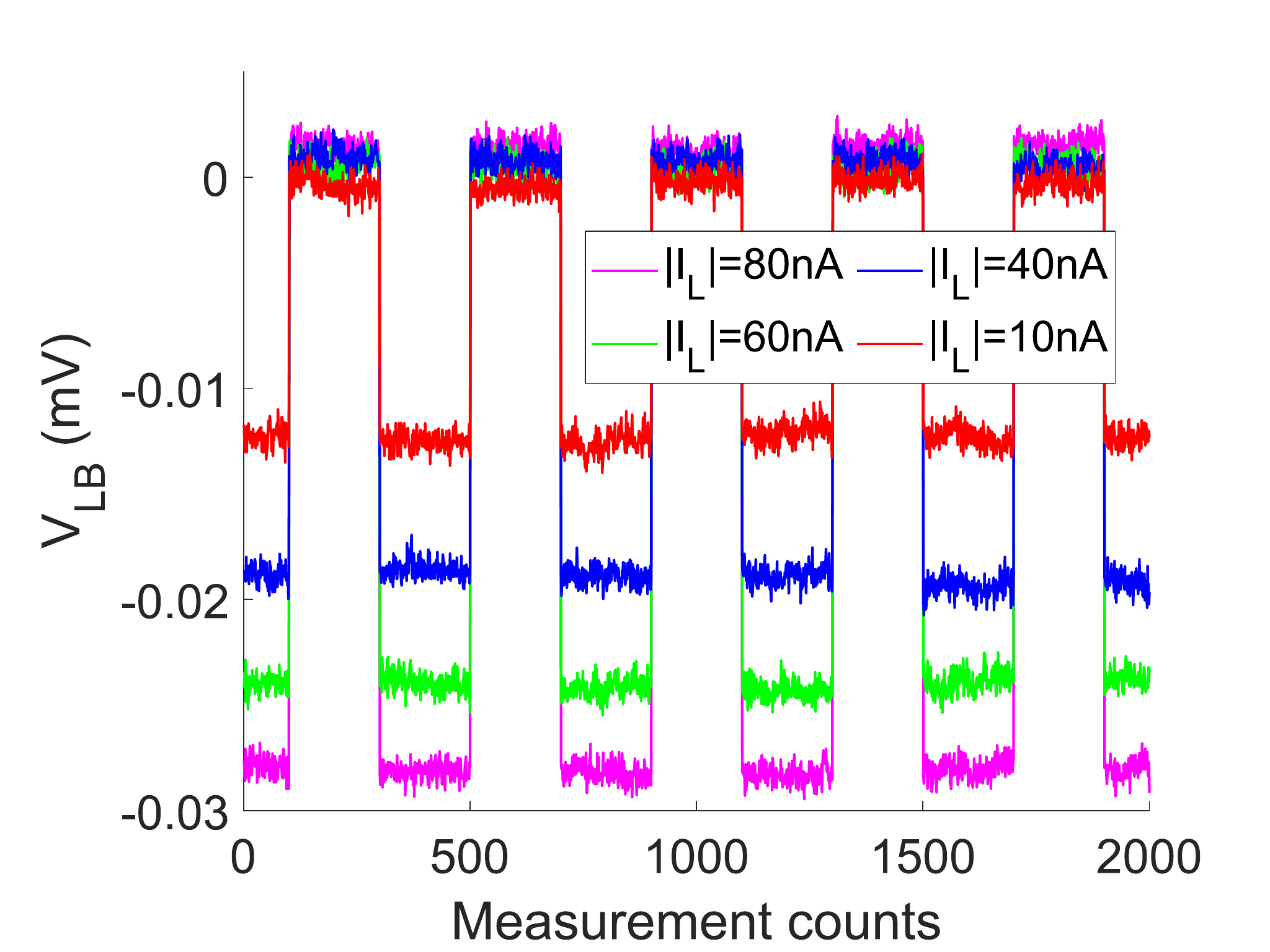}
    \caption {Rectification of square waves down to as low as 10 nA in amplitude. Here, the device operation is demonstrated at a finite field of 0.6 mT. $V_{G} = 0$, T = 1 K, and $I_{R} = -70$ nA.
    }
    \label{fig:FigS5}
\end{figure*}


\end{document}